\documentclass[fleqn,usenatbib]{mnras}

\usepackage{newtxtext,newtxmath}

\usepackage[T1]{fontenc}
\usepackage{ae,aecompl}

\usepackage{graphicx}	
\usepackage{amsmath}	
\usepackage{amssymb}	
\usepackage{siunitx}
\usepackage{subcaption}

\title[Causality distances]{Using variability and VLBI to measure cosmological distances}

\author[J. A. Hodgson]{Jeffrey A. Hodgson$^{1}$\thanks{Contact e-mail: \href{mailto:jhodgson@kasi.re.kr}{jhodgson@kasi.re.kr}}, Benjamin L'Huillier$^{2}$, Ioannis Liodakis$^{3}$, Sang-Sung Lee$^{1,4}$,\newauthor Arman Shafieloo$^{1,4}$
\\
$^{1}$Korea Astronomy and Space Science Institute, Daedeokdae-ro 776, Daejeon, South Korea \\
$^2$Yonsei University, 50 Yonsei-ro, Seodaemun-gu, Seoul 03722 Korea\\
$^{3}$Kavli Institute for Particle Astrophysics and Cosmology, Stanford University, 452 Lomita Mall, Stanford, CA 94305, USA \\
$^{4}$Korea University of Science and Technology, 217 Gajeong-ro, Yuseong-gu, Daejeon 34113, Korea }

\date{Accepted 2020 March 23. Received 2020 March 23; in original form 2019 October 25}

\pubyear{2020}

\begin{document}
\label{firstpage}
\pagerange{\pageref{firstpage}--\pageref{lastpage}}
\maketitle

\begin{abstract}
In this paper, we propose a new approach to determining cosmological distances to active galactic nuclei (AGN) via light travel-time arguments, which can be extended from nearby sources to very high redshift sources. The key assumption is that the variability seen in AGN is constrained by the speed of light and therefore provides an estimate of the linear size of an emitting region. This can then be compared with the angular size measured with very long baseline interferometry (VLBI) in order to derive a distance. We demonstrate this approach on a specific well studied low redshift ($z = 0.0178$) source 3C\,84 (NGC\,1275), which is the bright radio core of the Perseus Cluster. We derive an angular diameter distance including statistical errors of $D_{A} = 72^{+5}_{-6}$\,Mpc for this source, which is consistent with other distance measurements at this redshift. Possible sources of systematic errors and ways to correct for them are discussed.
\end{abstract}





\section{Introduction}
Independent measurements of distances and redshifts, $z$, allow astronomers to constrain cosmological models, since they both define the distance -- redshift relation.
When it was determined that type Ia supernovae (SNIa) could be standardised and therefore used to measure distances, this led to the discovery of the accelerated expansion of the Universe \citep{Riess1998,Perlmutter1999}.
The combination of SNIa \citep{2014A&A...568A..22B}, baryonic acoustic oscillations \citep{2005ApJ...633..560E,2017MNRAS.470.2617A}, and the cosmic microwave background \citep{2011ApJS..192...18K,planckH0} led to the emergence of the 
  concordance $\Lambda$CDM model, in which the energy density is dominated by dark energy as a cosmological constant $\Lambda$. 
In a flat Friedmann--Lema\^itre--Robertson--Walker Universe, the comoving distance is defined as
\begin{align}
    D(z) & =  \int_0^z 
    \frac{c\mathrm{d}z} {H(z)},\\
\intertext{where, in the $\Lambda$CDM model,}
H(z) & = H_0 \sqrt{\Omega_\mathrm{m} (1+z)^3 + 1-\Omega_\mathrm{m}} 
\end{align}
is the Hubble parameter, $H_{0}$ is the Hubble-Lema\^itre constant,  
and $\Omega_\mathrm{m}$ is the matter energy density at the current epoch.
The luminosity and angular diameter distances are defined as
\begin{align}
    D_\mathrm{L}(z) &= (1+z) D(z)\\
    \intertext{and}
    D_\mathrm{A}(z) &= \frac R \theta  = \frac{D(z)}{(1+z)}.
\end{align}
Type Ia supernovae can only be used up to redshift of around 2 \citep{jones13}, and there are tensions between direct local measurements of the Hubble-Lema\^itre constant and model-dependent estimates using Cosmic Microwave Background observations \citep{planckH0}. 
Therefore, independent distance measurements to extragalactic objects are desired. {We should emphasise here, that model independent distance indicators can have various important applications in physical cosmology. In particular, to test different aspects of cosmological models and theories of gravity. For instance, we can use these model independent measurements to test the FLRW metric \citep{clarkson2008, wiltshire2009, arman2010, ben2017, arman2018, qi2019, cao2019}, to test general relativity and some modified gravity models \citep{cao2012, arman2013b, cao15, cao2017, ben2018,arman2018, qi2017, xu2018, chen2019,  ben2019arxiv}, to test natural constants such as the speed of light \citep{cao2018}, to test cosmic duality relationships and to measure cosmic curvature \citep{arman2013a, qi2019, zheng2020arxiv, cao2019b, qi2019b}. Using a combination of such model independent distance indicators can also be used to measure some key cosmological parameters such as the Hubble Constant \citep{suyu2017, liao2015, jee2019, liao2019arman, wong2019, liao2020arxiv}. }
Amongst the most energetic objects in the Universe are active galactic nuclei (AGN). 
AGN are the nuclei of massive galaxies that sometimes produce relativistic jets of material launched from near a central super-massive black hole (SMBH). When these jets are not aligned close to our line-of-sight, AGN are observed as radio galaxies, whereas if the jet is aligned to within a small angle to our line-of-sight, they are observed as blazars \citep{urry95}. 
Blazars are amongst the most consistently bright objects in the Universe and can be observed at redshifts as high as $\sim 7$ \citep{mortlock11}.
Attempts have been made to measure distances to AGN in various ways, with some claiming deviations from the expected cosmology at high redshifts \citep{risalti17,risaliti19,turner19}. VLBI has also been used to attempt to measure cosmological distances. The approach pioneered by \citet{gurvits98} attempted to measure cosmological parameters by assuming that AGN could be used as a standardisable rod. \citet{vishwakarma01} used this dataset and compared it with supernovae data, and found it was not possible to differentiate different cosmological models with the VLBI data of \citet{gurvits98}. \citet{cao15} revisited this technique and investigated the evolution of the standard rod by assuming a Planck cosmology. \citet{cau17a,cau17b} then introduced a cosmology independent method for calibrating the standard rod and was able to provide reasonable constraints on cosmological parameters. Our approach differs from this by using the speed of light to normalise the rod. This approach was first attempted by \citet{wiik2001}, which found that that the apparent angular sizes of AGN maximised at $z\sim2$.
In this paper, we demonstrate the method on the famous nearby radio galaxy 3C\,84 (NGC\,1275) and discuss possible systematic errors.
The source is known to exhibit extremely high energy emission despite not exhibiting strong relativistic effects \citep{jor17,yannis18}, and has multiple independent measures of distance \citep{2007A&A...465...71T,2009ApJ...700.1097H}, thus making it an ideal source to test our methodology.

\begin{figure*}
    \centering
    \includegraphics[width=0.94\textwidth]{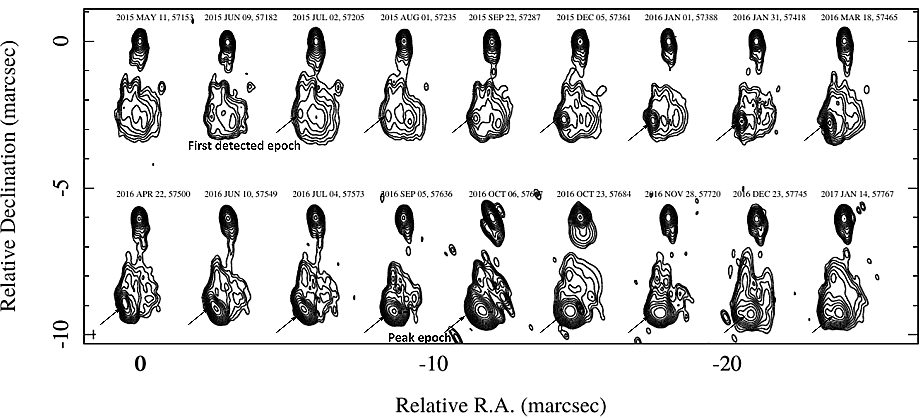}
    \caption{43 GHz (7\,mm) VLBI maps of 3C 84 from May 11 2015 (MJD 57153) until Jan 14 2017 (MJD 57767). Contours: -1, 1, 2, 4, 8, 16, 32, 64\% of peak flux density. The emitting region used for the distance measurement is pointed out with a black arrow. The size and flux density is determined by Gaussian model-fitting the emitting region directly. The results of this fitting are found in Table \ref{modfits}. The flux densities obtained are used in the light-curve shown in Fig. \ref{fig:LCplot}.  }
    \label{fig:3c84maps}
\end{figure*}

\section{Methods}\label{methods}
The core assumption of the method is that the angular radius of a source as measured via VLBI ($\theta_{\mathrm{VLBI}}$, measured in milliarcseconds) is equivalent to the linear radius ($R$, measured in meters) inferred by causality arguments. Therefore:
\begin{equation}
    R = \frac{c \Delta t}{(1+z)},
\end{equation}
where $\Delta t$ is the variability (or light-crossing) timescale. The observed (i.e. angular) size of this region on the sky depends on the distance. Thus, the inferred angular radius from the light-crossing time is:
\begin{equation}
    \theta_{\rm var} = \frac{R}{D_{\rm A}} = \frac{c \Delta t }{(1+z)D_{\rm A}}.
\end{equation}
Hence, if we measure the size directly using VLBI, we can set $\theta_{var}=\theta_{VLBI}$ and solve for the angular diameter distance:
\begin{equation}
    D_{\rm A} = \frac{c \Delta t }{\theta_{\mathrm{VLBI}}(1+z)}.
\end{equation}

\section{Observations}
We obtained publicly available
high resolution maps of the source at 7\,mm (43 GHz) observing wavelength from the Boston University blazar monitoring (VLBA-BU-BLAZAR) program \citep{jor05,jor17}. In practice, VLBI observations are measuring the incomplete Fourier transform of the sky brightness distribution, from which an image is produced using the CLEAN algorithm \citep{clean} and phase and amplitude self-calibration. In order to parameterise features within these images, we fitted elliptical or circular Gaussian models directly to the interferometric visibilities, providing us with angular size and flux density measurements of emission regions within the map. We performed this analysis using standard routines in the program DIFMAP \citep{difmap}. To ensure amplitude calibration accuracy, the total VLBI flux densities were compared against total intensity measurements of the source and corrected accordingly if needed \citep[e.g.,][]{kim19}. The maps are shown in Fig. \ref{fig:3c84maps}. We use these model-fitted flux densities as the measurements from which the light-curve shown in Fig. \ref{fig:LCplot} is derived. The flux density, radius of the major axis of the full-width-half-maximum (FWHM) of the fitted Gaussians and the beam sizes is presented in Table \ref{modfits}. As the variability time-scale is a 1-dimensional quantity and the model-fits are 2-dimensional, we must choose an axis to compare the size against. We consider the major axis to be the most conservative approach. This is nevertheless an assumption, which we discuss further in Section \ref{systematics}. In order to convert the FWHM of the Gaussian to a more realistic spherical or thin-disk geometry, we multiply the Gaussian by either a factor of 1.6 or 1.8 respectively \citep{marscher77}. Since we are unsure of the true geometry, we use a compromise scaling factor of 1.7 and include the ambiguity in the error analysis. Because the source is at a low redshift of $z=0.0178$  \citep{strauss92}, peculiar velocities can introduce systematic errors \citep{zwicky99,davis11}. We have introduced a conservative 10\% error \citep{hudson97} on the redshift to account for this.

\section{Results}

\begin{figure}
    \centering
    \includegraphics[width=\linewidth]{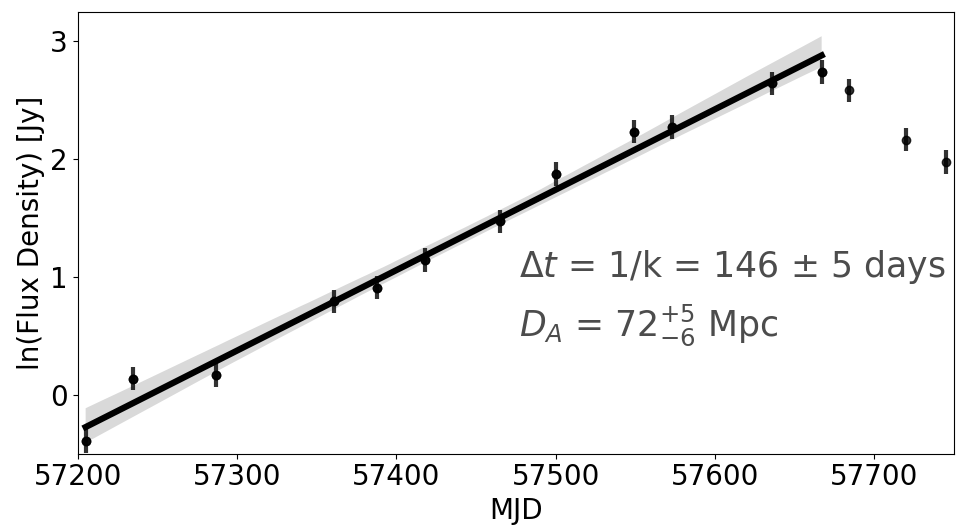}
    \caption{The light-curve of the emitting region, as determined from directly model-fitting the maps shown in Fig. \ref{fig:3c84maps}. The variability timescale is defined as the reciprocal of the slope, $k$, fit to the log of the flux densities between the minimum and maximum flux densities observed during a flare in the source. The solid black line is the best fit to the slope and the gray shaded area is the 95\% confidence interval. Assuming that this variability is reasonably constrained by the speed of light, this provides an estimate of the linear size of the emitting region. This is compared with the angular radius of the emitting region measured using VLBI ($\theta_{\mathrm{VLBI}} = 0.20 \pm 0.02$\,mas) in order to determine a angular diameter distance of $D_{A} = 72^{+5}_{-6}$\,Mpc.}
    \label{fig:LCplot}
\end{figure}

The VLBI morphology of the source is currently dominated by two main emitting regions: the region thought to be near the central SMBH, which is the northernmost bright emission region in the maps shown in Fig. \ref{fig:3c84maps}, and a slowly moving emission feature to the south and which has been studied recently by several authors \citep{nagai16,hiura18,hodgson18}. The slowly moving emission feature had a large flare occur in it, beginning in mid 2015, which is pointed out with black arrows \citep{hodgson18}.
It should be noted that the relevant quantity is not the relative motion of the emission region from the SMBH but its flux density and size. 
We use this flare and directly model-fit it (acquiring the size and flux density information of the emitting region directly) to perform our distance measurements. The critical epochs are the beginning and peak of a flare. We discuss this in the next paragraph.
In order to determine the variability (light-crossing, $\Delta t$) timescale, we fit the slope, $k$, to the logarithmic flux density (ln $S$) as a  function of time (modified Julian date, MJD) in a range of flux density from $S_{\rm min}$ to $S_{\rm max}$ (see Fig. \ref{fig:LCplot}). We determined $S_{\rm min}$ to be the first epoch in which the emission region was reliably detected in the VLBI images. See Appendix \ref{var_discussion} for a more in-depth discussion. The variability timescale is then $1/k$ (i.e. the e-folding timescale of the flare), which is a standard method in the literature \citep{terasranta94,valtaoja,jor05,hovatta09,jor17}. We determined a variability timescale of $\Delta t = 145 \pm 5$ days. The radius of the FWHM fitted to the emitting region at the peak of the flare is measured to be $\theta_{\mathrm{VLBI}} = 0.20 \pm 0.02$\,mas, with the component being easily resolved by the interferometer. We measure the size at the peak of the flare. This is because if one imagines a photon travelling across the source at the speed of light, it would necessarily be at least restricted to the size that we measure at the peak, since a photon would have had to travel at least that distance in order to make the size that we measure.
\subsection{Error analysis}
Errors were propagated using a Monte Carlo approach by creating a normal distribution for each observable. The mean of the distribution was set as the observed value and the standard deviation of the distribution was set as the error on the observed value. A distribution made of 10 000 samples was made for each variable and these distributions were used in the place of the variables presented in the equations shown in this paper. The 1$\sigma$ final errors were determined by finding the 68\% limits of the final distribution. This leads to an estimate for the angular diameter distance of $D_{A} = 72^{+5}_{-6}$\,Mpc (corresponding to a Hubble constant of $H_{0}=73^{+5}_{-6}$ km/s/Mpc).

\subsection{Sources of systematic error}\label{systematics}

A major source of systematic errors can come from the observations themselves. A limitation of the results presented here is that we are highly cadence and resolution limited using existing telescopes and monitoring programs. With these data, which were observed with a roughly monthly cadence. It is possible that there are flares which are shorter than a month in duration, that are missed due to the limited cadence of the observations. Similarly, we may not measure the correct size due to not observing exactly at the peak of a flare. VLBI flux density calibration can be somewhat uncertain and include flux-scaling errors. This can require comparison with total-intensity measurements \citep[e.g.][]{kim19} which could also lead to systematic errors. While angular resolution is not a problem with these observations, if we observe at high redshift, it could be possible that the source becomes unresolved. In this case, we could place limits on the size of the emitting region by using the major and minor axes of the observing beam \citep{gurvits98,cao15}. Furthermore, the results obtained here were achieved at an observing frequency of 43 GHz. According to AGN jet models \citep[e.g.][]{bk79,bloom1996}, the variability time-scale is expected to vary as a function of frequency.  We, therefore, suggest continued observations at both higher and lower frequencies to further verify our methods.
Nevertheless, even with perfect observations, there are several other potential sources of systematic errors. They include i) the assumption that the variability is constrained by the speed of light; ii) uncertainties in the geometry of the emission region; and iii) determining when a flare begins and ends.
i) The key assumption of the method is that the observed variability is reasonably constrained by the speed of light. On a physical level, the emission from 3C 84 is due to synchrotron radiation by electrons (or other charged particles) being accelerated around magnetic field lines travelling at nearly the speed of light. Given the physics of the radiation, we believe it likely that the emission is tightly constrained by the speed of light, but not exactly. This assumption has been indirectly investigated by \citet{liodakis15} and \citet{yannis18}. In these studies, they investigated different methods for determining the Doppler factor in a large sample of AGN. They found that the variability Doppler factor - which depends on the causality assumption - best-fit the population. Implicitly this suggests that the causality assumption is valid. However, the results of \citet{liodakis15} are model dependent, because it assumes a source distribution model. Additionally, a way 
to directly test the assumption would be to use this method on microquasars with known parallax distances \citep[e.g.][]{reid14}. We intend to perform these observations in the future.
ii) There is also some uncertainty regarding the geometry of the emission region. In this proof-of-concept paper, we are unable to differentiate between a spherical geometry, a thin disk geometry or non-face-on orientations of thin disk geometries or more complex geometries \citep[e.g.,][]{protheroe2002}. However with careful analysis of the visibilities of sufficiently high resolution observations, it should be possible to determine the true source geometry \citep{pearson95}. Furthermore, we assume that the variability timescale equates to the radius of the emitting region, and that we are sensitive to the longest axis of a project ellipsoid. Given that the emitting regions are likely shock-fronts, we believe this to be a reasonable assumption. However with a careful analysis of visibilities in simple sources, this can be accounted for or potentially also modelled. 
iii) Determining the variability timescale is a critical parameter in deriving distances, with the critical parameters being when a flare begins and ends. This is discussed in Appendix \ref{var_discussion}. Some potential ways to correct for uncertainties in the variability timescale could be to use $\gamma$-ray flaring as a proxy for a flare onset or using polarisation measurements.
In the case of 3C 84, we are nevertheless confident that we are reasonably accurately measuring the variability timescale, as the distance we derive is consistent with other methods. In particular, the presence of a type Ia supernova in the galaxy allows a distance measurement of $62.5$ to $82.8$ Mpc \citep{2009ApJ...700.1097H}, while  
Tully-Fisher measurements to the brightest galaxy yield a distance of $53.9$ to $68.9$ Mpc \citep{2007A&A...465...71T}.
We should emphasise that systematic errors such as these should not have any redshift dependence. This means that these errors can affect the absolute scaling of the distance measurements, but not the shape as a function redshift. Therefore, this can affect measurements of the Hubble Constant, but should not affect measurements of the energy content of the universe. However, quasars and blazars exhibit relativistic effects that must be taken into account in order to determine accurate distances at the highest redshifts. In an upcoming paper, we will investigate other sources and demonstrate how these relativistic effects can be accounted for and therefore applied to a larger range of sources, and potentially bridging the gap between supernova and CMB measurements. However, for strongly relativistic sources, there could be redshift dependent systematic errors. This could arise from a selection bias, where we preferentially select only the most relativistic sources at the highest redshifts. How an effect like this would manifest itself in practice is not yet clear. For radio galaxies which are only mildly relativistic, we can apply our method directly.

\section{Conclusions}

We have presented a proof-of-concept measurement of an angular diameter distance that is independent of cosmological model assumptions and of the distance ladder. It is worth noting that this method can also be applied to non-AGN type sources. In order for our method to work, a source need only have its flux density variability reasonably approximated by the speed of light and be resolvable by our instruments.
In order to perform these observations, cadence and high resolution monitoring will be required. Within this context, there are currently plans to convert the Mopra telescope in Australia to be compatible with the quasi-optics of the Korean VLBI Network. The KVN is capable of observing at four frequencies simultaneously \citep{lee_kvn,hodgson2016}, allowing us to confirm that the variability timescales and sizes change as a function of frequency. We will be able to perform multi-frequency, high-cadence and high-resolution monitoring of AGN over a large range of redshifts, allowing us to constrain both the Hubble constant and the matter density of the Universe. \\
\newline

\begin{table}
	\centering
	\caption{Table of Gaussian model-fitted values used in this paper.}
	\label{modfits}
	\begin{tabular}{ccccc} 
		\hline
		MJD & Flux density & Half FWHM & Beam (maj x min, PA) \\
		 &  [Jy] & [mas] & [mas x mas], [$^{\circ}$] \\
		\hline
		 57205 &   0.67 $\pm$  0.07 &  0.16 $\pm$   0.01 & 0.160 x 0.337,5.27  \\
        57235 &   1.14 $\pm$  0.15 &  0.29 $\pm$   0.02 & 0.154 x 0.280,7.87 \\
        57287 &   1.18 $\pm$  0.18 &  0.07 $\pm$   0.01 & 0.166 x 0.311,3.49 \\
        57361 &   2.21 $\pm$  0.22 &  0.11 $\pm$   0.01 & 0.169 x 0.276,6.01  \\
        57388 &   2.48 $\pm$  0.25 &  0.08 $\pm$   0.01 & 0.168 x 0.327,7.48 \\
        57418 &   3.14 $\pm$  0.31 &  0.08 $\pm$   0.01 & 0.158 x 0.281,4.73  \\
        57465 &   4.36 $\pm$  0.44 &  0.07 $\pm$   0.01 & 0.162 x 0.314,3.16 \\
        57500 &  6.53 $\pm$  0.65 &  0.18 $\pm$   0.02 & 0.158 x 0.278,5.67 \\
        57549 &  9.33 $\pm$  0.93 &  0.22 $\pm$   0.02 & 0.164 x 0.314,15.90  \\
        57573 &  9.73 $\pm$  0.97 &  0.20 $\pm$   0.02 & 0.161 x 0.302,13.91  \\
        57636 &  14.06 $\pm$  1.44 &  0.18 $\pm$   0.01 & 0.160 x 0.312,1.50 \\
        57667 &  15.68 $\pm$  1.53 &  0.20 $\pm$   0.02 & 0.185 x 0.423,24.24 \\
        57684 &  13.25 $\pm$  1.64 &  0.21 $\pm$   0.02 & 0.161 x 0.295,13.97 \\
        57720 &   8.73 $\pm$  0.87 &  0.23 $\pm$   0.02 & 0.161 x 0.300,5.75 \\
        57745 &   7.21 $\pm$  0.72 &  0.20 $\pm$   0.02 & 0.160 x 0.289,7.82 \\

		\hline
	\end{tabular}
\end{table}

\appendix

\section{Determining the variability timescale}\label{var_discussion}

Accurately determining the variability timescale ($\Delta t$) is of critical importance for determining the distance, with it being sensitive to determining the onset of the flare. In this case, we consider the flare to begin in the first epoch that the flaring component was reliably detected in the VLBI images. Furthermore, a very bright $\gamma$-ray flare has been associated with the emergence of this component \citep{hodgson18}, with it peaking in 2015.81 (MJD 57318), although its onset is approximately 2015.6 -- 2015.7 (MJD 57240 - 57280), which is consistent with the emergence of the component. Nevertheless, we can explore how the distance measurement is affected by different flare definitions. In Fig. \ref{fig:LCplot}, we can see that the flux density slightly decreases at $\sim$MJD 57300. If we select this epoch as when the flare begins, we derive a distance of $D_{A}=70^{+6}_{-6}$\,Mpc, which makes a $\sim$1\% difference and is still consistent with distances measured using other methods \citep{2009ApJ...700.1097H,2007A&A...465...71T}. We plan to fully investigate the most appropriate way to determine the variability timescales in our upcoming project.

\textbf{Acknowledgements:} The author would like to acknowledge the help of Alan Marscher and Svetlana Jorstad for their help in preparing this manuscript and providing the data for 3C 84. This work by Jeffrey A. Hodgson was supported by Korea Research Fellowship Program through the National Research Foundation of Korea(NRF) funded by the Ministry of Science and ICT(2018H1D3A1A02032824). This study makes use of 43 GHz VLBA data from the VLBA-BU Blazar Monitoring Program (VLBA-BU-BLAZAR; \url{http://www.bu.edu/blazars/VLBAproject.html}), funded by NASA through Fermi Guest Investigator grant 80NSSC17K0649. The VLBA is an instrument of the National Radio Astronomy Observatory. BL would like to acknowledge the support of the National Research Foundation of Korea (NRF-2019R1I1A1A01063740). This work was supported by the Samsung Science and Technology Foundation under Project Number SSTF-BA1801-04. 

\bibliographystyle{mnras}
\bibliography{biblio}

\begin{thebibliography}{}
\makeatletter
\relax
\def\mn@urlcharsother{\let\do\@makeother \do\$\do\&\do\#\do\^\do\_\do\%\do\~}
\def\mn@doi{\begingroup\mn@urlcharsother \@ifnextchar [ {\mn@doi@}
  {\mn@doi@[]}}
\def\mn@doi@[#1]#2{\def\@tempa{#1}\ifx\@tempa\@empty \href
  {http://dx.doi.org/#2} {doi:#2}\else \href {http://dx.doi.org/#2} {#1}\fi
  \endgroup}
\def\mn@eprint#1#2{\mn@eprint@#1:#2::\@nil}
\def\mn@eprint@arXiv#1{\href {http://arxiv.org/abs/#1} {{\tt arXiv:#1}}}
\def\mn@eprint@dblp#1{\href {http://dblp.uni-trier.de/rec/bibtex/#1.xml}
  {dblp:#1}}
\def\mn@eprint@#1:#2:#3:#4\@nil{\def\@tempa {#1}\def\@tempb {#2}\def\@tempc
  {#3}\ifx \@tempc \@empty \let \@tempc \@tempb \let \@tempb \@tempa \fi \ifx
  \@tempb \@empty \def\@tempb {arXiv}\fi \@ifundefined
  {mn@eprint@\@tempb}{\@tempb:\@tempc}{\expandafter \expandafter \csname
  mn@eprint@\@tempb\endcsname \expandafter{\@tempc}}}

\bibitem[\protect\citeauthoryear{{Alam} et~al.,}{{Alam}
  et~al.}{2017}]{2017MNRAS.470.2617A}
{Alam} S.,  et~al., 2017, \mn@doi [\mnras] {10.1093/mnras/stx721}, \href
  {http://adsabs.harvard.edu/abs/2017MNRAS.470.2617A} {470, 2617}

\bibitem[\protect\citeauthoryear{{Betoule} et~al.,}{{Betoule}
  et~al.}{2014}]{2014A&A...568A..22B}
{Betoule} M.,  et~al., 2014, \mn@doi [\aap] {10.1051/0004-6361/201423413},
  \href {http://adsabs.harvard.edu/abs/2014A\%26A...568A..22B} {568, A22}

\bibitem[\protect\citeauthoryear{{Blandford} \& {K{\"o}nigl}}{{Blandford} \&
  {K{\"o}nigl}}{1979}]{bk79}
{Blandford} R.~D.,  {K{\"o}nigl} A.,  1979, \mn@doi [\apj] {10.1086/157262},
  \href {https://ui.adsabs.harvard.edu/abs/1979ApJ...232...34B} {232, 34}

\bibitem[\protect\citeauthoryear{{Bloom} \& {Marscher}}{{Bloom} \&
  {Marscher}}{1996}]{bloom1996}
{Bloom} S.~D.,  {Marscher} A.~P.,  1996, \mn@doi [\apj] {10.1086/177092}, \href
  {https://ui.adsabs.harvard.edu/abs/1996ApJ...461..657B} {461, 657}

\bibitem[\protect\citeauthoryear{{Cao}, {Pan}, {Biesiada}, {Godlowski}  \&
  {Zhu}}{{Cao} et~al.}{2012}]{cao2012}
{Cao} S.,  {Pan} Y.,  {Biesiada} M.,  {Godlowski} W.,   {Zhu} Z.-H.,  2012,
  \mn@doi [\jcap] {10.1088/1475-7516/2012/03/016}, \href
  {https://ui.adsabs.harvard.edu/abs/2012JCAP...03..016C} {2012, 016}

\bibitem[\protect\citeauthoryear{{Cao}, {Biesiada}, {Zheng}  \& {Zhu}}{{Cao}
  et~al.}{2015}]{cao15}
{Cao} S.,  {Biesiada} M.,  {Zheng} X.,   {Zhu} Z.-H.,  2015, \mn@doi [\apj]
  {10.1088/0004-637X/806/1/66}, \href
  {https://ui.adsabs.harvard.edu/abs/2015ApJ...806...66C} {806, 66}

\bibitem[\protect\citeauthoryear{{Cao}, {Zheng}, {Biesiada}, {Qi}, {Chen}  \&
  {Zhu}}{{Cao} et~al.}{2017a}]{cau17a}
{Cao} S.,  {Zheng} X.,  {Biesiada} M.,  {Qi} J.,  {Chen} Y.,   {Zhu} Z.-H.,
  2017a, \mn@doi [\aap] {10.1051/0004-6361/201730551}, \href
  {https://ui.adsabs.harvard.edu/abs/2017A&A...606A..15C} {606, A15}

\bibitem[\protect\citeauthoryear{{Cao}, {Li}, {Biesiada}, {Xu}, {Cai}  \&
  {Zhu}}{{Cao} et~al.}{2017b}]{cao2017}
{Cao} S.,  {Li} X.,  {Biesiada} M.,  {Xu} T.,  {Cai} Y.,   {Zhu} Z.-H.,  2017b,
  \mn@doi [\apj] {10.3847/1538-4357/835/1/92}, \href
  {https://ui.adsabs.harvard.edu/abs/2017ApJ...835...92C} {835, 92}

\bibitem[\protect\citeauthoryear{{Cao}, {Biesiada}, {Jackson}, {Zheng}, {Zhao}
  \& {Zhu}}{{Cao} et~al.}{2017c}]{cau17b}
{Cao} S.,  {Biesiada} M.,  {Jackson} J.,  {Zheng} X.,  {Zhao} Y.,   {Zhu}
  Z.-H.,  2017c, \mn@doi [\jcap] {10.1088/1475-7516/2017/02/012}, \href
  {https://ui.adsabs.harvard.edu/abs/2017JCAP...02..012C} {2017, 012}

\bibitem[\protect\citeauthoryear{{Cao}, {Qi}, {Biesiada}, {Zheng}, {Xu}  \&
  {Zhu}}{{Cao} et~al.}{2018}]{cao2018}
{Cao} S.,  {Qi} J.,  {Biesiada} M.,  {Zheng} X.,  {Xu} T.,   {Zhu} Z.-H.,
  2018, \mn@doi [\apj] {10.3847/1538-4357/aae5f7}, \href
  {https://ui.adsabs.harvard.edu/abs/2018ApJ...867...50C} {867, 50}

\bibitem[\protect\citeauthoryear{{Cao}, {Qi}, {Cao}, {Biesiada}, {Li}, {Pan}
  \& {Zhu}}{{Cao} et~al.}{2019a}]{cao2019}
{Cao} S.,  {Qi} J.,  {Cao} Z.,  {Biesiada} M.,  {Li} J.,  {Pan} Y.,   {Zhu}
  Z.-H.,  2019a, \mn@doi [Scientific Reports] {10.1038/s41598-019-47616-4},
  \href {https://ui.adsabs.harvard.edu/abs/2019NatSR...911608C} {9, 11608}

\bibitem[\protect\citeauthoryear{{Cao}, {Qi}, {Biesiada}, {Zheng}, {Xu}, {Pan}
  \& {Zhu}}{{Cao} et~al.}{2019b}]{cao2019b}
{Cao} S.,  {Qi} J.,  {Biesiada} M.,  {Zheng} X.,  {Xu} T.,  {Pan} Y.,   {Zhu}
  Z.-H.,  2019b, \mn@doi [Physics of the Dark Universe]
  {10.1016/j.dark.2019.100274}, \href
  {https://ui.adsabs.harvard.edu/abs/2019PDU....24..274C} {24, 100274}

\bibitem[\protect\citeauthoryear{{Chen}, {Sesana}  \& {Conselice}}{{Chen}
  et~al.}{2019}]{chen2019}
{Chen} S.,  {Sesana} A.,   {Conselice} C.~J.,  2019, \mn@doi [\mnras]
  {10.1093/mnras/stz1722}, \href
  {https://ui.adsabs.harvard.edu/abs/2019MNRAS.488..401C} {488, 401}

\bibitem[\protect\citeauthoryear{{Clarkson}, {Bassett}  \& {Lu}}{{Clarkson}
  et~al.}{2008}]{clarkson2008}
{Clarkson} C.,  {Bassett} B.,   {Lu} T. H.-C.,  2008, \mn@doi [\prl]
  {10.1103/PhysRevLett.101.011301}, \href
  {https://ui.adsabs.harvard.edu/abs/2008PhRvL.101a1301C} {101, 011301}

\bibitem[\protect\citeauthoryear{{Davis} et~al.,}{{Davis}
  et~al.}{2011}]{davis11}
{Davis} T.~M.,  et~al., 2011, \mn@doi [\apj] {10.1088/0004-637X/741/1/67},
  \href {https://ui.adsabs.harvard.edu/abs/2011ApJ...741...67D} {741, 67}

\bibitem[\protect\citeauthoryear{{Eisenstein} et~al.,}{{Eisenstein}
  et~al.}{2005}]{2005ApJ...633..560E}
{Eisenstein} D.~J.,  et~al., 2005, \mn@doi [\apj] {10.1086/466512}, \href
  {http://adsabs.harvard.edu/abs/2005ApJ...633..560E} {633, 560}

\bibitem[\protect\citeauthoryear{{Falco} et~al.,}{{Falco}
  et~al.}{1999}]{zwicky99}
{Falco} E.~E.,  et~al., 1999, \mn@doi [\pasp] {10.1086/316343}, \href
  {http://adsabs.harvard.edu/abs/1999PASP..111..438F} {111, 438}

\bibitem[\protect\citeauthoryear{{Gurvits}, {Kellermann}  \& {Frey}}{{Gurvits}
  et~al.}{1999}]{gurvits98}
{Gurvits} L.~I.,  {Kellermann} K.~I.,   {Frey} S.,  1999, \aap, \href
  {https://ui.adsabs.harvard.edu/abs/1999A&A...342..378G} {342, 378}

\bibitem[\protect\citeauthoryear{{Hicken}, {Wood-Vasey}, {Blondin}, {Challis},
  {Jha}, {Kelly}, {Rest}  \& {Kirshner}}{{Hicken}
  et~al.}{2009}]{2009ApJ...700.1097H}
{Hicken} M.,  {Wood-Vasey} W.~M.,  {Blondin} S.,  {Challis} P.,  {Jha} S.,
  {Kelly} P.~L.,  {Rest} A.,   {Kirshner} R.~P.,  2009, \mn@doi [\apj]
  {10.1088/0004-637X/700/2/1097}, \href
  {http://adsabs.harvard.edu/abs/2009ApJ...700.1097H} {700, 1097}

\bibitem[\protect\citeauthoryear{{Hiura} et~al.,}{{Hiura}
  et~al.}{2018}]{hiura18}
{Hiura} K.,  et~al., 2018, \mn@doi [\pasj] {10.1093/pasj/psy078}, \href
  {http://ads.nao.ac.jp/abs/2018PASJ...70...83H} {70, 83}

\bibitem[\protect\citeauthoryear{{Hodgson}, {Lee}, {Zhao}, {Algaba}, {Yun},
  {Jung}  \& {Byun}}{{Hodgson} et~al.}{2016}]{hodgson2016}
{Hodgson} J.~A.,  {Lee} S.-S.,  {Zhao} G.-Y.,  {Algaba} J.-C.,  {Yun} Y.,
  {Jung} T.,   {Byun} D.-Y.,  2016, \mn@doi [Journal of Korean Astronomical
  Society] {10.5303/JKAS.2016.49.4.137}, \href
  {https://ui.adsabs.harvard.edu/abs/2016JKAS...49..137H} {49, 137}

\bibitem[\protect\citeauthoryear{{Hodgson} et~al.,}{{Hodgson}
  et~al.}{2018}]{hodgson18}
{Hodgson} J.~A.,  et~al., 2018, \mn@doi [\mnras] {10.1093/mnras/stx3041}, \href
  {http://ads.nao.ac.jp/abs/2018MNRAS.475..368H} {475, 368}

\bibitem[\protect\citeauthoryear{{H{\"o}gbom}}{{H{\"o}gbom}}{1974}]{clean}
{H{\"o}gbom} J.~A.,  1974, \aaps, \href
  {https://ui.adsabs.harvard.edu/abs/1974A%26AS...15..417H} {15, 417}

\bibitem[\protect\citeauthoryear{{Hovatta}, {Valtaoja}, {Tornikoski}  \&
  {L{\"a}hteenm{\"a}ki}}{{Hovatta} et~al.}{2009}]{hovatta09}
{Hovatta} T.,  {Valtaoja} E.,  {Tornikoski} M.,   {L{\"a}hteenm{\"a}ki} A.,
  2009, \mn@doi [\aap] {10.1051/0004-6361:200811150}, \href
  {https://ui.adsabs.harvard.edu/abs/2009A&A...494..527H} {494, 527}

\bibitem[\protect\citeauthoryear{{Hudson}, {Lucey}, {Smith}  \&
  {Steel}}{{Hudson} et~al.}{1997}]{hudson97}
{Hudson} M.~J.,  {Lucey} J.~R.,  {Smith} R.~J.,   {Steel} J.,  1997, \mn@doi
  [\mnras] {10.1093/mnras/291.3.488}, \href
  {http://adsabs.harvard.edu/abs/1997MNRAS.291..488H} {291, 488}

\bibitem[\protect\citeauthoryear{{Jee}, {Suyu}, {Komatsu}, {Fassnacht},
  {Hilbert}  \& {Koopmans}}{{Jee} et~al.}{2019}]{jee2019}
{Jee} I.,  {Suyu} S.~H.,  {Komatsu} E.,  {Fassnacht} C.~D.,  {Hilbert} S.,
  {Koopmans} L. V.~E.,  2019, \mn@doi [Science] {10.1126/science.aat7371},
  \href {https://ui.adsabs.harvard.edu/abs/2019Sci...365.1134J} {365, 1134}

\bibitem[\protect\citeauthoryear{{Jones} et~al.,}{{Jones}
  et~al.}{2013}]{jones13}
{Jones} D.~O.,  et~al., 2013, \mn@doi [\apj] {10.1088/0004-637X/768/2/166},
  \href {http://adsabs.harvard.edu/abs/2013ApJ...768..166J} {768, 166}

\bibitem[\protect\citeauthoryear{{Jorstad} et~al.,}{{Jorstad}
  et~al.}{2005}]{jor05}
{Jorstad} S.~G.,  et~al., 2005, \mn@doi [\aj] {10.1086/444593}, \href
  {http://ads.nao.ac.jp/abs/2005AJ....130.1418J} {130, 1418}

\bibitem[\protect\citeauthoryear{{Jorstad} et~al.,}{{Jorstad}
  et~al.}{2017}]{jor17}
{Jorstad} S.~G.,  et~al., 2017, \mn@doi [\apj] {10.3847/1538-4357/aa8407},
  \href {http://ads.nao.ac.jp/abs/2017ApJ...846...98J} {846, 98}

\bibitem[\protect\citeauthoryear{{Kim} et~al.,}{{Kim} et~al.}{2019}]{kim19}
{Kim} J.-Y.,  et~al., 2019, \mn@doi [\aap] {10.1051/0004-6361/201832920}, \href
  {http://ads.nao.ac.jp/abs/2019A%26A...622A.196K} {622, A196}

\bibitem[\protect\citeauthoryear{{Komatsu} et~al.,}{{Komatsu}
  et~al.}{2011}]{2011ApJS..192...18K}
{Komatsu} E.,  et~al., 2011, \mn@doi [\apjs] {10.1088/0067-0049/192/2/18},
  \href {http://cdsads.u-strasbg.fr/abs/2011ApJS..192...18K} {192, 18}

\bibitem[\protect\citeauthoryear{{L'Huillier} \& {Shafieloo}}{{L'Huillier} \&
  {Shafieloo}}{2017}]{ben2017}
{L'Huillier} B.,  {Shafieloo} A.,  2017, \mn@doi [\jcap]
  {10.1088/1475-7516/2017/01/015}, \href
  {https://ui.adsabs.harvard.edu/abs/2017JCAP...01..015L} {2017, 015}

\bibitem[\protect\citeauthoryear{{L'Huillier}, {Shafieloo}  \&
  {Kim}}{{L'Huillier} et~al.}{2018}]{ben2018}
{L'Huillier} B.,  {Shafieloo} A.,   {Kim} H.,  2018, \mn@doi [\mnras]
  {10.1093/mnras/sty398}, \href
  {https://ui.adsabs.harvard.edu/abs/2018MNRAS.476.3263L} {476, 3263}

\bibitem[\protect\citeauthoryear{{L'Huillier}, {Shafieloo}, {Polarski}  \&
  {Starobinsky}}{{L'Huillier} et~al.}{2020}]{ben2019arxiv}
{L'Huillier} B.,  {Shafieloo} A.,  {Polarski} D.,   {Starobinsky} A.~A.,  2020,
  \mn@doi [\mnras\ in press] {10.1093/mnras/staa633}, \href
  {https://ui.adsabs.harvard.edu/abs/2019arXiv190605991L} {p. arXiv:1906.05991}

\bibitem[\protect\citeauthoryear{{Lee} et~al.,}{{Lee} et~al.}{2014}]{lee_kvn}
{Lee} S.-S.,  et~al., 2014, \mn@doi [\aj] {10.1088/0004-6256/147/4/77}, \href
  {https://ui.adsabs.harvard.edu/abs/2014AJ....147...77L} {147, 77}

\bibitem[\protect\citeauthoryear{{Liao} et~al.,}{{Liao}
  et~al.}{2015}]{liao2015}
{Liao} K.,  et~al., 2015, \mn@doi [\apj] {10.1088/0004-637X/800/1/11}, \href
  {https://ui.adsabs.harvard.edu/abs/2015ApJ...800...11L} {800, 11}

\bibitem[\protect\citeauthoryear{{Liao}, {Shafieloo}, {Keeley}  \&
  {Linder}}{{Liao} et~al.}{2019a}]{liao2019arman}
{Liao} K.,  {Shafieloo} A.,  {Keeley} R.~E.,   {Linder} E.~V.,  2019a, \mn@doi
  [\apjl] {10.3847/2041-8213/ab5308}, \href
  {https://ui.adsabs.harvard.edu/abs/2019ApJ...886L..23L} {886, L23}

\bibitem[\protect\citeauthoryear{{Liao}, {Shafieloo}, {Keeley}  \&
  {Linder}}{{Liao} et~al.}{2019b}]{wong2019}
{Liao} K.,  {Shafieloo} A.,  {Keeley} R.~E.,   {Linder} E.~V.,  2019b, \mn@doi
  [\apjl] {10.3847/2041-8213/ab5308}, \href
  {https://ui.adsabs.harvard.edu/abs/2019ApJ...886L..23L} {886, L23}

\bibitem[\protect\citeauthoryear{{Liao}, {Shafieloo}, {Keeley}  \&
  {Linder}}{{Liao} et~al.}{2020}]{liao2020arxiv}
{Liao} K.,  {Shafieloo} A.,  {Keeley} R.~E.,   {Linder} E.~V.,  2020, arXiv
  e-prints, \href {https://ui.adsabs.harvard.edu/abs/2020arXiv200210605L} {p.
  arXiv:2002.10605}

\bibitem[\protect\citeauthoryear{{Liodakis} \& {Pavlidou}}{{Liodakis} \&
  {Pavlidou}}{2015}]{liodakis15}
{Liodakis} I.,  {Pavlidou} V.,  2015, \mn@doi [\mnras] {10.1093/mnras/stv2028},
  \href {https://ui.adsabs.harvard.edu/abs/2015MNRAS.454.1767L} {454, 1767}

\bibitem[\protect\citeauthoryear{{Liodakis}, {Hovatta}, {Huppenkothen},
  {Kiehlmann}, {Max-Moerbeck}  \& {Readhead}}{{Liodakis}
  et~al.}{2018}]{yannis18}
{Liodakis} I.,  {Hovatta} T.,  {Huppenkothen} D.,  {Kiehlmann} S.,
  {Max-Moerbeck} W.,   {Readhead} A. C.~S.,  2018, \mn@doi [\apj]
  {10.3847/1538-4357/aae2b7}, \href
  {https://ui.adsabs.harvard.edu/\#abs/2018ApJ...866..137L} {866, 137}

\bibitem[\protect\citeauthoryear{{Marscher}}{{Marscher}}{1977}]{marscher77}
{Marscher} A.~P.,  1977, \mn@doi [\apj] {10.1086/155467}, \href
  {http://adsabs.harvard.edu/abs/1977ApJ...216..244M} {216, 244}

\bibitem[\protect\citeauthoryear{{Mortlock} et~al.,}{{Mortlock}
  et~al.}{2011}]{mortlock11}
{Mortlock} D.~J.,  et~al., 2011, \mn@doi [\nat] {10.1038/nature10159}, \href
  {https://ui.adsabs.harvard.edu/\#abs/2011Natur.474..616M} {474, 616}

\bibitem[\protect\citeauthoryear{{Nagai}, {Chida}, {Kino}, {Orienti},
  {D'Ammando}, {Giovannini}  \& {Hiura}}{{Nagai} et~al.}{2016}]{nagai16}
{Nagai} H.,  {Chida} H.,  {Kino} M.,  {Orienti} M.,  {D'Ammando} F.,
  {Giovannini} G.,   {Hiura} K.,  2016, \mn@doi [Astronomische Nachrichten]
  {10.1002/asna.201512267}, \href
  {http://ads.nao.ac.jp/abs/2016AN....337...69N} {337, 69}

\bibitem[\protect\citeauthoryear{{Pearson}}{{Pearson}}{1995}]{pearson95}
{Pearson} T.~J.,  1995, in {Zensus} J.~A.,  {Diamond} P.~J.,   {Napier} P.~J.,
  eds,  Astronomical Society of the Pacific Conference Series Vol. 82, Very
  Long Baseline Interferometry and the VLBA. p.~267

\bibitem[\protect\citeauthoryear{{Perlmutter} et~al.,}{{Perlmutter}
  et~al.}{1999}]{Perlmutter1999}
{Perlmutter} S.,  et~al., 1999, \mn@doi [\apj] {10.1086/307221}, \href
  {https://ui.adsabs.harvard.edu/abs/1999ApJ...517..565P} {517, 565}

\bibitem[\protect\citeauthoryear{{Planck Collaboration} et~al.,}{{Planck
  Collaboration} et~al.}{2018}]{planckH0}
{Planck Collaboration} et~al., 2018, arXiv e-prints, \href
  {http://ads.nao.ac.jp/abs/2018arXiv180706209P} {p. arXiv:1807.06209}

\bibitem[\protect\citeauthoryear{{Protheroe}}{{Protheroe}}{2002}]{protheroe2002}
{Protheroe} R.~J.,  2002, \mn@doi [\pasa] {10.1071/AS02008}, \href
  {https://ui.adsabs.harvard.edu/abs/2002PASA...19..486P} {19, 486}

\bibitem[\protect\citeauthoryear{{Qi}, {Cao}, {Biesiada}, {Zheng}  \&
  {Zhu}}{{Qi} et~al.}{2017}]{qi2017}
{Qi} J.-Z.,  {Cao} S.,  {Biesiada} M.,  {Zheng} X.,   {Zhu} Z.-H.,  2017,
  \mn@doi [European Physical Journal C] {10.1140/epjc/s10052-017-5069-1}, \href
  {https://ui.adsabs.harvard.edu/abs/2017EPJC...77..502Q} {77, 502}

\bibitem[\protect\citeauthoryear{{Qi}, {Cao}, {Biesiada}, {Zheng}, {Ding}  \&
  {Zhu}}{{Qi} et~al.}{2019a}]{qi2019}
{Qi} J.,  {Cao} S.,  {Biesiada} M.,  {Zheng} X.,  {Ding} X.,   {Zhu} Z.-H.,
  2019a, \mn@doi [\prd] {10.1103/PhysRevD.100.023530}, \href
  {https://ui.adsabs.harvard.edu/abs/2019PhRvD.100b3530Q} {100, 023530}

\bibitem[\protect\citeauthoryear{{Qi}, {Cao}, {Zhang}, {Biesiada}, {Wu}  \&
  {Zhu}}{{Qi} et~al.}{2019b}]{qi2019b}
{Qi} J.-Z.,  {Cao} S.,  {Zhang} S.,  {Biesiada} M.,  {Wu} Y.,   {Zhu} Z.-H.,
  2019b, \mn@doi [\mnras] {10.1093/mnras/sty3175}, \href
  {https://ui.adsabs.harvard.edu/abs/2019MNRAS.483.1104Q} {483, 1104}

\bibitem[\protect\citeauthoryear{{Reid}, {McClintock}, {Steiner}, {Steeghs},
  {Remillard}, {Dhawan}  \& {Narayan}}{{Reid} et~al.}{2014}]{reid14}
{Reid} M.~J.,  {McClintock} J.~E.,  {Steiner} J.~F.,  {Steeghs} D.,
  {Remillard} R.~A.,  {Dhawan} V.,   {Narayan} R.,  2014, \mn@doi [\apj]
  {10.1088/0004-637X/796/1/2}, \href
  {https://ui.adsabs.harvard.edu/\#abs/2014ApJ...796....2R} {796, 2}

\bibitem[\protect\citeauthoryear{{Riess} et~al.,}{{Riess}
  et~al.}{1998}]{Riess1998}
{Riess} A.~G.,  et~al., 1998, \mn@doi [\aj] {10.1086/300499}, \href
  {http://adsabs.harvard.edu/abs/1998AJ....116.1009R} {116, 1009}

\bibitem[\protect\citeauthoryear{{Risaliti} \& {Lusso}}{{Risaliti} \&
  {Lusso}}{2017}]{risalti17}
{Risaliti} G.,  {Lusso} E.,  2017, \mn@doi [Astronomische Nachrichten]
  {10.1002/asna.201713351}, \href
  {http://adsabs.harvard.edu/abs/2017AN....338..329R} {338, 329}

\bibitem[\protect\citeauthoryear{{Risaliti} \& {Lusso}}{{Risaliti} \&
  {Lusso}}{2019}]{risaliti19}
{Risaliti} G.,  {Lusso} E.,  2019, \mn@doi [Nature Astronomy]
  {10.1038/s41550-018-0657-z}, \href
  {http://adsabs.harvard.edu/abs/2019NatAs.tmp..195R} {}

\bibitem[\protect\citeauthoryear{{Shafieloo} \& {Clarkson}}{{Shafieloo} \&
  {Clarkson}}{2010}]{arman2010}
{Shafieloo} A.,  {Clarkson} C.,  2010, \mn@doi [\prd]
  {10.1103/PhysRevD.81.083537}, \href
  {https://ui.adsabs.harvard.edu/abs/2010PhRvD..81h3537S} {81, 083537}

\bibitem[\protect\citeauthoryear{{Shafieloo}, {Kim}  \& {Linder}}{{Shafieloo}
  et~al.}{2013a}]{arman2013b}
{Shafieloo} A.,  {Kim} A.~G.,   {Linder} E.~V.,  2013a, \mn@doi [\prd]
  {10.1103/PhysRevD.87.023520}, \href
  {https://ui.adsabs.harvard.edu/abs/2013PhRvD..87b3520S} {87, 023520}

\bibitem[\protect\citeauthoryear{{Shafieloo}, {Majumdar}, {Sahni}  \&
  {Starobinsky}}{{Shafieloo} et~al.}{2013b}]{arman2013a}
{Shafieloo} A.,  {Majumdar} S.,  {Sahni} V.,   {Starobinsky} A.~A.,  2013b,
  \mn@doi [\jcap] {10.1088/1475-7516/2013/04/042}, \href
  {https://ui.adsabs.harvard.edu/abs/2013JCAP...04..042S} {2013, 042}

\bibitem[\protect\citeauthoryear{{Shafieloo}, {L'Huillier}  \&
  {Starobinsky}}{{Shafieloo} et~al.}{2018}]{arman2018}
{Shafieloo} A.,  {L'Huillier} B.,   {Starobinsky} A.~A.,  2018, \mn@doi [\prd]
  {10.1103/PhysRevD.98.083526}, \href
  {https://ui.adsabs.harvard.edu/abs/2018PhRvD..98h3526S} {98, 083526}

\bibitem[\protect\citeauthoryear{{Shepherd}}{{Shepherd}}{1997}]{difmap}
{Shepherd} M.~C.,  1997, in {Hunt} G.,  {Payne} H.,  eds,  Astronomical Society
  of the Pacific Conference Series Vol. 125, Astronomical Data Analysis
  Software and Systems VI. p.~77

\bibitem[\protect\citeauthoryear{{Strauss}, {Huchra}, {Davis}, {Yahil},
  {Fisher}  \& {Tonry}}{{Strauss} et~al.}{1992a}]{3c84_z}
{Strauss} M.~A.,  {Huchra} J.~P.,  {Davis} M.,  {Yahil} A.,  {Fisher} K.~B.,
  {Tonry} J.,  1992a, \mn@doi [\apjs] {10.1086/191730}, \href
  {http://adsabs.harvard.edu/abs/1992ApJS...83...29S} {83, 29}

\bibitem[\protect\citeauthoryear{{Strauss}, {Huchra}, {Davis}, {Yahil},
  {Fisher}  \& {Tonry}}{{Strauss} et~al.}{1992b}]{strauss92}
{Strauss} M.~A.,  {Huchra} J.~P.,  {Davis} M.,  {Yahil} A.,  {Fisher} K.~B.,
  {Tonry} J.,  1992b, \mn@doi [\apjs] {10.1086/191730}, \href
  {http://adsabs.harvard.edu/abs/1992ApJS...83...29S} {83, 29}

\bibitem[\protect\citeauthoryear{{Suyu} et~al.,}{{Suyu}
  et~al.}{2017}]{suyu2017}
{Suyu} S.~H.,  et~al., 2017, \mn@doi [\mnras] {10.1093/mnras/stx483}, \href
  {https://ui.adsabs.harvard.edu/abs/2017MNRAS.468.2590S} {468, 2590}

\bibitem[\protect\citeauthoryear{{Terasranta} \& {Valtaoja}}{{Terasranta} \&
  {Valtaoja}}{1994}]{terasranta94}
{Terasranta} H.,  {Valtaoja} E.,  1994, \aap, \href
  {https://ui.adsabs.harvard.edu/abs/1994A&A...283...51T} {283, 51}

\bibitem[\protect\citeauthoryear{{Theureau}, {Hanski}, {Coudreau}, {Hallet}  \&
  {Martin}}{{Theureau} et~al.}{2007}]{2007A&A...465...71T}
{Theureau} G.,  {Hanski} M.~O.,  {Coudreau} N.,  {Hallet} N.,   {Martin} J.-M.,
   2007, \mn@doi [\aap] {10.1051/0004-6361:20066187}, \href
  {http://adsabs.harvard.edu/abs/2007A%26A...465...71T} {465, 71}

\bibitem[\protect\citeauthoryear{{Turner} \& {Shabala}}{{Turner} \&
  {Shabala}}{2019}]{turner19}
{Turner} R.,  {Shabala} S.,  2019, arXiv e-prints, \href
  {https://ui.adsabs.harvard.edu/\#abs/2019arXiv190312308T} {p.
  arXiv:1903.12308}

\bibitem[\protect\citeauthoryear{{Urry} \& {Padovani}}{{Urry} \&
  {Padovani}}{1995}]{urry95}
{Urry} C.~M.,  {Padovani} P.,  1995, \mn@doi [\pasp] {10.1086/133630}, \href
  {http://adsabs.harvard.edu/abs/1995PASP..107..803U} {107, 803}

\bibitem[\protect\citeauthoryear{{Valtaoja}, {L{\"a}hteenm{\"a}ki},
  {Ter{\"a}sranta}  \& {Lainela}}{{Valtaoja} et~al.}{1999}]{valtaoja}
{Valtaoja} E.,  {L{\"a}hteenm{\"a}ki} A.,  {Ter{\"a}sranta} H.,   {Lainela} M.,
   1999, \mn@doi [\apjs] {10.1086/313170}, \href
  {https://ui.adsabs.harvard.edu/abs/1999ApJS..120...95V} {120, 95}

\bibitem[\protect\citeauthoryear{{Vishwakarma}}{{Vishwakarma}}{2001}]{vishwakarma01}
{Vishwakarma} R.~G.,  2001, \mn@doi [Classical and Quantum Gravity]
  {10.1088/0264-9381/18/7/301}, \href
  {https://ui.adsabs.harvard.edu/abs/2001CQGra..18.1159V} {18, 1159}

\bibitem[\protect\citeauthoryear{{Wiik} \& {Valtaoja}}{{Wiik} \&
  {Valtaoja}}{2001}]{wiik2001}
{Wiik} K.,  {Valtaoja} E.,  2001, \mn@doi [\aap] {10.1051/0004-6361:20000285},
  \href {https://ui.adsabs.harvard.edu/abs/2001A&A...366.1061W} {366, 1061}

\bibitem[\protect\citeauthoryear{{Wiltshire}}{{Wiltshire}}{2009}]{wiltshire2009}
{Wiltshire} D.~L.,  2009, \mn@doi [\prd] {10.1103/PhysRevD.80.123512}, \href
  {https://ui.adsabs.harvard.edu/abs/2009PhRvD..80l3512W} {80, 123512}

\bibitem[\protect\citeauthoryear{{Xu}, {Cao}, {Qi}, {Biesiada}, {Zheng}  \&
  {Zhu}}{{Xu} et~al.}{2018}]{xu2018}
{Xu} T.,  {Cao} S.,  {Qi} J.,  {Biesiada} M.,  {Zheng} X.,   {Zhu} Z.-H.,
  2018, \mn@doi [\jcap] {10.1088/1475-7516/2018/06/042}, \href
  {https://ui.adsabs.harvard.edu/abs/2018JCAP...06..042X} {2018, 042}

\bibitem[\protect\citeauthoryear{{Zheng}, {Liao}, {Biesiada}, {Cao}, {Liu}  \&
  {Zhu}}{{Zheng} et~al.}{2020}]{zheng2020arxiv}
{Zheng} X.,  {Liao} K.,  {Biesiada} M.,  {Cao} S.,  {Liu} T.-H.,   {Zhu} Z.-H.,
   2020, arXiv e-prints, \href
  {https://ui.adsabs.harvard.edu/abs/2020arXiv200209909Z} {p. arXiv:2002.09909}

\makeatother
\end{thebibliography}

\bsp	
\label{lastpage}
\end{document}